# Chemical Mechanical Planarization for Ta-based Superconducting Quantum Devices

Ekta Bhatia[1,a], Soumen Kar[1], Jakub Nalaskowski[1], Tuan Vo[1], Stephen Olson[1], Hunter Frost[2], John Mucci[1], Brian Martinick[1], Pui Yee Hung[1], Ilyssa Wells[1], Sandra Schujman[1], Satyavolu S. Papa Rao[1,2,b]

[1]NY CREATES, Albany, NY 12203, USA
[2]College of Nanoscale Science and Engineering, SUNY Polytechnic Institute, Albany, NY 12203, USA

a) Electronic mail: ebhatia@sunypoly.edu
b) Electronic mail: spaparao@sunypoly.edu

We report on the development of a chemical mechanical planarization (CMP) process for thick damascene Ta structures with pattern feature sizes down to 100 nm. This CMP process is the core of the fabrication sequence for scalable superconducting integrated circuits at 300 mm wafer scale. This work has established the elements of the various CMP-related design rules that can be followed by a designer for the layout of circuits that include Ta-based coplanar waveguide resonators, capacitors, and interconnects for tantalum-based qubits and single flux quantum (SFQ) circuits. The fabrication of these structures utilizes 193 nm optical lithography, along with 300 mm process tools for dielectric deposition, reactive ion etch, wet-clean, CMP and in-line metrology, all tools typical for a 300 mm wafer CMOS foundry. Process development was guided by measurements of physical and electrical characteristics of the planarized structures. Physical characterization such as atomic force microscopy across the 300 mm wafer surface showed local topography was less than 5 nm. Electrical characterization confirmed low leakage at room temperature, and less than 12% within wafer sheet resistance variation, for damascene Ta line-widths ranging from 100 nm to 3 µm. Run-to-run reproducibility was also evaluated. Effects of process integration choices including deposited thickness of Ta are discussed.

## I.  INTRODUCTION

Tantalum is a promising superconducting material for improving the coherence time of superconducting qubits when used to form the capacitor and microwave resonators in superconducting circuits [1]. Recently, two research groups [2, 3] have reported high coherence times in the range of 0.1 to 0.5 ms for transmon qubits made on sapphire substrates using α-Ta (BCC phase) for wiring. The higher coherence time in both the reports was ascribed to the difference in native oxides present on Ta *vs.* other metals (Nb, Al). However, in order to enable the fabrication of Ta-based qubits in a truly scalable process, it would be advantageous to develop processing on 300 mm Si substrates (which are commonly used for IC fabrication), and move away from smaller diameter sapphire substrates. This will need new processes to be developed while retaining the advantages provided by Ta for qubit coherence times. Creating chips on 300 mm wafers with qubits that are interconnected with Ta lines offers a scaling path for superconducting quantum computing.

In addition to its advantages for superconducting quantum computing, Ta is a possible candidate to replace Nb for applications in certain SFQ circuits. For SFQ circuits that are coupled to quantum circuits, and operate at temperatures below 1 K, the lower superconducting transition temperature of Ta *vs.* Nb is not an issue. Ta has other advantages over Nb: lower sensitivity to hydrogen than Nb [4] and lower solid solubility of oxygen [5]. The functioning of SFQ circuits depends on the self and mutual inductances of interconnects but such inductances are sensitive to fabrication process variation and to the specifics of chip layout. Hence, fabrication of SFQ circuits with better process control than demonstrated previously, and leveraging a well-established design-rule manual can open the door to very large scale integrated SFQ circuits. Ta is a widely used material in the CMOS fab, thus processes, and consumables developed by the CMOS IC industry over the past two decades can be advantageously leveraged and modified to suit the purposes of superconducting circuits. For these reasons, it is feasible, and timely, to develop a controllable process for fabricating superconducting Ta interconnects, with predictable characteristics that can be relied upon by the circuit designer.

In this paper, we demonstrate for the first time the design of a chemical mechanical planarization (CMP) process to create damascene Ta patterns with linewidths ranging from 100 nm to 3 μm, with a post-CMP nanowire thickness of 80 – 85 nm. We have established an optimal CMP process for Ta interconnect patterns, with good within-wafer uniformity and wafer-to-wafer repeatability, supported by electrical measurements and physical characterization including X-ray fluorescence (XRF), atomic force microscopy (AFM), scanning electron microscopy (SEM), and optical microscopy. We have established the basic elements of CMP-related design rules for Ta interconnects that can be utilized for both quantum computing and SFQ circuit applications. Designs using CMP require the incorporation of layout elements called cheese and dummy fill [6]. The process margin associated with these layout elements was studied to establish the parameters within which good CMP performance can be expected.

## II. EXPERIMENTAL

### A. Process Flow Considerations

In this section, the process sequence is described, and the desiderata for superconducting circuits using Ta are laid out. An $SiO_2$/SiN dielectric bilayer on a Si (100) wafer is patterned using 193 nm optical lithography followed by a reactive ion etch (RIE) process that creates trenches stopping on silicon, as shown in figure 1(a). The exposed silicon in the trenches is etched to deepen the trench into silicon, with sloped sidewalls that expose (111) planes in silicon, using a hydroxide-based etch. The sloped sidewalls help with filling sputtered metal into the trenches, and avoiding the formation of seams that would be present with vertical trench sidewalls. The SiN hard mask was removed using hot phosphoric acid etch, as shown in figure 1(b). The SiN hard-mask over the $SiO_2$ allows for the full thickness of the oxide to be retained during the trench patterning process. The SiN can also be used for other patterns to be created on the wafer surface (for example, deep alignment marks for multi-layer Ta interconnects) which are not used in this work. Following the removal of SiN, the metallic films required for the interconnect (ALD TaN/Ta) are deposited on the surface, as shown in figure 1(c).

While Ta is an element widely used in the IC industry (as a part of the liner for Cu interconnects), the α phase of Ta is specifically necessary for superconducting logic and quantum applications. The α phase of Ta has a superconducting transition temperature of ~3.2 K to 4.3 K depending on deposition conditions [7, 8]. The formation of α-Ta is promoted by variety of underlayers [9]; a good underlayer choice is ALD TaN [10]. Without the underlying layer, sputter deposition of Ta on oxide forms the β phase which has an extremely low superconducting temperature (~mK) [11]. The formation of α-Ta can be confirmed at room temperature by x-ray diffraction as well as by resistivity measurements since it has about 7x lower resistivity than the β phase [13,14].

CMP of the metallized wafer results in metal-filled trenches, as illustrated in figure 1(d). The oxide hardmask is only partly removed during the CMP – with the exact amount remaining being a function of the CMP process being developed.

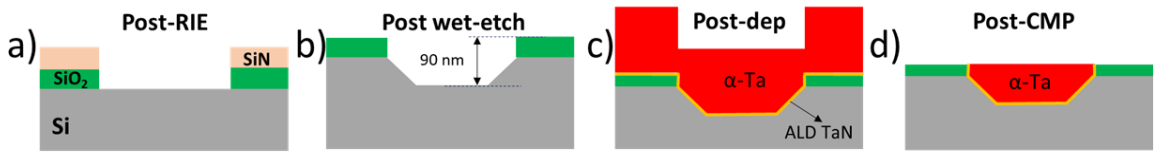

FIG. 1. Different steps in the fabrication scheme: a) Schematic showing wafer cross-section after $SiO_2$/SiN RIE (dimensions not to scale). b) Etching exposes silicon in trenches using hydroxide based etch, followed by removal of SiN. c) Schematic showing wafer cross-section after deposition of ALD TaN liner/Ta. d) Schematic showing wafer cross-section after CMP.

The thickness of Ta after CMP is completed should be greater than its penetration depth in order to limit dissipative losses when used in superconducting circuits such as SFQ circuits and superconducting resonators [15]. The surface of the Ta film roughly duplicates the topography when it is sputtered onto a surface with patterned trenches. Since the ratio of removal rates at the low areas to the removal rate at high areas is not zero, it is necessary to give the CMP process enough incoming thickness to accomplish planarization. The specific overburden that is required for a given incoming topography is dependent on slurry and pad specifics. As a rule of thumb, the deposited thickness should be at least 1.5x, and preferably around 2x nm for incoming topography of x nm. In the case of Ta, film stress created during the deposition process precludes the deposition of a very thick film, hence the starting thickness for CMP also needs to be chosen with care [16].

An optimized CMP process leaves the Ta surface locally smooth and the wafer surface planar (within a few nanometers), along with good uniformity from center to edge [17]. Such a planar surface makes the deposition and patterning of the next layer of interconnect easier, and more controllable, with associated advantages in SFQ circuit performance. Metal CMP relies on having as uniform a pattern as possible across the wafer surface, since the CMP process acts simultaneously on all the materials exposed, each of which necessarily have a different local CMP rate. This is particularly important after the field overburden is cleared, and Ta, ALD TaN and $SiO_2$ are all simultaneously exposed to the slurry and pad. Therefore, variations in local areal density of Ta features will impact the CMP process locally. From the process engineer's perspective, a uniform Ta pattern density is desired (irrespective of its specific value), where Ta pattern density, or metal density, is typically defined as the fraction of the area occupied by Ta patterns [6]. In contrast, the designer would appreciate the flexibility afforded by permitting any value of local Ta pattern density. The compromise that has been used in the IC industry, and which we have adopted, is to choose the average pattern density to be as close to 50% as possible.

Dummy patterns are automatically placed in areas without designed Ta features in order that the average pattern density is closer to 50%. The use of dummy patterns is called 'dummy fill'. As a result of uniform local metal pattern density (and post CMP metal thickness), the self-inductance of the conductors is more uniform and predictable (both within a chip and from design to design). This will help with the modeling of inductances and improving the ease of circuit design in SFQ circuits. Due to these considerations, we chose a damascene metal CMP approach to fabricate the patterned structures with dimensions varying from 100 nm to 3 µm, in preference to patterning metal with a RIE process.

## B. Process Control Upstream of Ta CMP

### 1. Dimensional Uniformity

Line-widths were measured on 300 mm wafers, at various points during the fabrication sequence, using a Critical Dimension Scanning Electron Microscope (CD-SEM), equipped with automated pattern recognition capability. Measurements made across the wafer surface after lithography, RIE, and CMP are summarized in figure 2. Figure 2(a) shows the variation of linewidth as a function of radial position on three wafers at the three steps that affect pattern width. The within-wafer non-uniformity of photoresist line width is <2%. The uniformity worsens after RIE for wafer radii higher than 75 mm, while still retaining wafer-to-wafer repeatability. The lot-to-lot variability is <2% for each process step, based on the data collected for 22 lots (73 wafers) over a period of 8 months, as shown in figure 2(b). The term "lot" refers to a batch of identical wafers or ICs that are processed at the same time and under the same conditions, carrying chips of a specific version of a device. The point corresponding to each lot presents the wafer average of post-RIE CD for all the wafers that ran in that lot. Figure 2(d) shows the result of a topography measurement after anisotropic Si etch. The scan was performed along the black line shown in the image of the AFM feature (which includes line widths varying from 100 nm to 5 µm) shown in figure 2(c). The trench depth of the 100 nm wide line was not resolvable due to the limitation of AFM tip size but all other line widths were resolvable and confirmed the trench depth of 110 nm, as shown in figure 2(d). Trench depth was the same (within measurement error of ±5 nm) across the wafer, between wafers in the same lot, and from lot-to-lot (based on data from wafers in 5 different lots, processed over a period of 4 months).

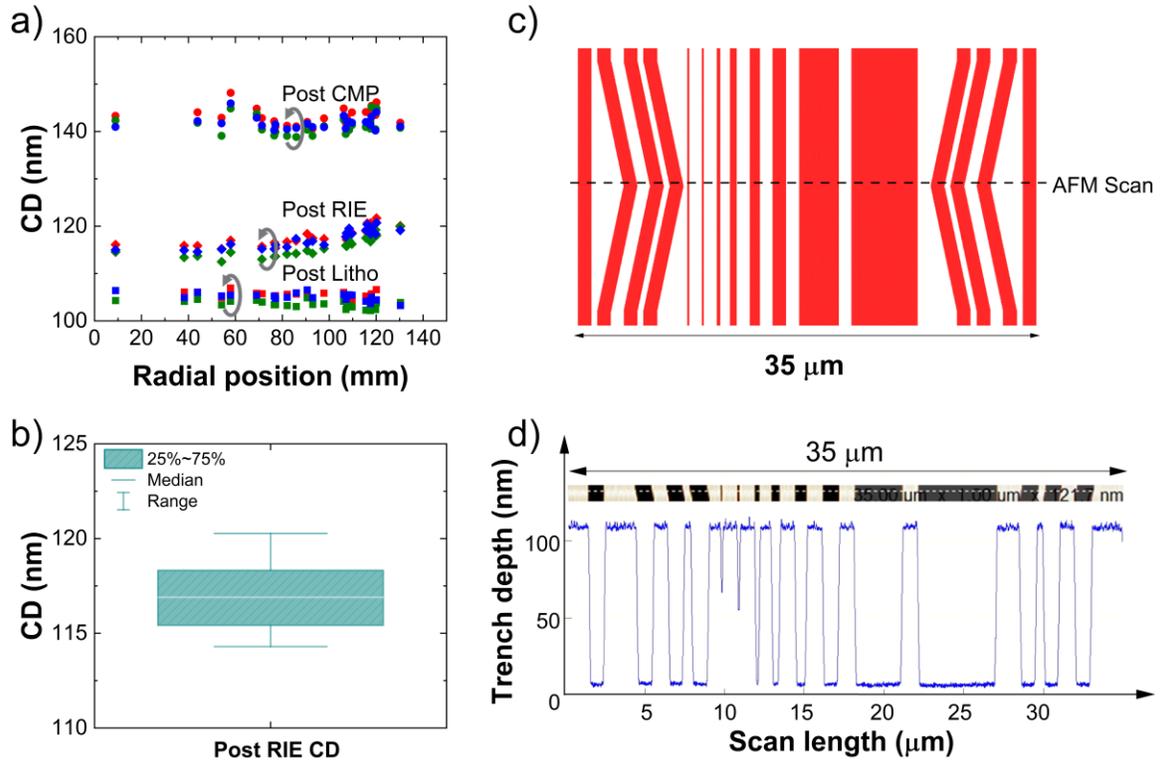

FIG. 2. (a) Across wafer CD uniformity for 110 nm wide (nominal) feature after lithography, RIE, and CMP. (b) Box and whisker plot showing lot to lot variability of post-RIE CD with the center line representing the mean value. (c) Layout image of AFM test feature with trenches in red color and $SiO_2$/SiN hard mask in white space in between with trench widths varying from 100 nm to 5 μm. (d) AFM scan across the AFM test feature confirming the 110 nm trench depth after hydroxide based etch.

## 2. α-Ta deposition

α-Ta films were deposited at 300 mm scale using magnetron sputtering in a multi-chamber sputter deposition tool. Characterization of this film was done with a bilayer film on silicon as shown in figure 3(a). A 3 nm thick underlayer of ALD TaN film, deposited using atomic layer deposition with a commercially available Ta precursor, serves two purposes: Firstly, it acts as a reaction barrier between Ta and Si [12]. Secondly, it helps promote the formation of α-Ta [10].

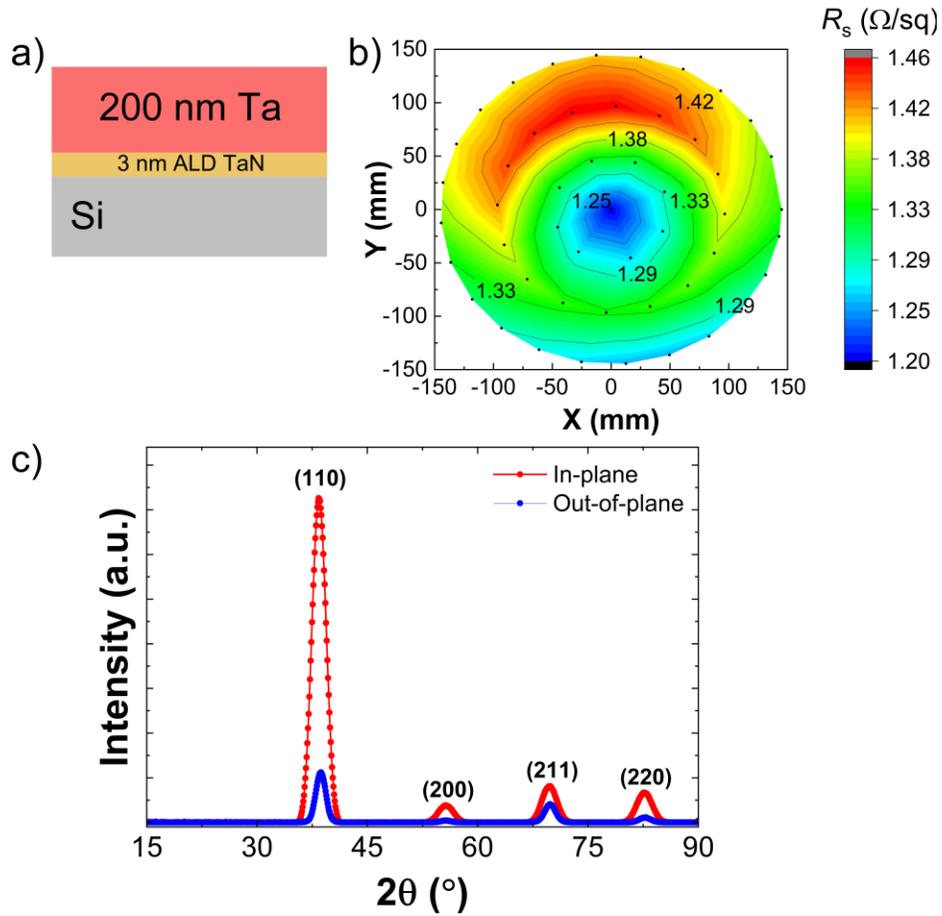

FIG. 3. (a) Schematic cross-section of the blanket films to obtain α-Ta. (b) Sheet-resistance map over the entire wafer at room temperature. (c) In-plane and out-of-plane XRD confirming the BCC phase of α–Ta film.

Figure 3(b) shows the sheet resistance map obtained from 49 points distributed across the wafer surface with an edge exclusion of 3 mm. The calculated within-wafer resistance non-uniformity (one standard deviation as a percentage of the median resistance) is better than 5%. The resistivity is 23 µΩ-cm, calculated from the resistance and XRR thickness measurement, providing one confirmation that Ta is present in α-phase [13,14]. Figure 3(c) shows the in-plane and out-of-plane XRD, providing the second confirmation, that Ta has the body-centered cubic crystal structure expected of α-Ta, with a lattice constant of 3.3 Å. Table I lists all the film properties of blanket Ta films at room temperature.

TABLE I. List of film properties of α-Ta blanket film at room temperature.

| *Property* | Value |
|---|---|
| Film stress (MPa) | 681 |
| Resistivity (μΩ-cm) | 23 |
| Crystal structure | BCC |
| Deposition rate (nm/s) | 0.73 |
| Thickness non-uniformity | 4.9% |
| Sheet-resistance non-uniformity | 4.9% |

## 3. Ta CMP

In most 300 mm CMP processes, a combination of hard pad and soft pad is typically used since it permits the desirable aspects of both pads to be leveraged such as the higher polish rate, better planarization, and in-situ process endpoint enabled by the hard pad with the low defectivity of the soft pad, when used as the last step of a multi-pad CMP process [18]. The CMP process design considerations include dividing the process between the hard pad and soft pad, and the choice of slurry to be used on each polish pad [18]. It should be noted that commercial slurries are designed with many goals in mind, including removal rate, the ratio of removal rate between various materials that are to be planarized, static etch minimization, corrosion protection and overall low defectivity. These design considerations influence the choice of the abrasive particle, as well as chemistry package (oxidizer, chelating agents, surfactants, and corrosion inhibitors). These formulation specifics are closely held, and not discussed in open literature. In this work, a commercially available alumina-based slurry was used for the hard-pad planarization step, with a commercially available silica-based slurry for the soft-pad buff step. Both are provided by their suppliers to the CMOS industry as slurries that can accomplish Ta removal.

After the removal of SiN with hot phosphoric acid, the step height of the trench is correspondingly decreased, from 110 nm to a value of 90 nm (figure 1(b)). The preferred thickness of incoming Ta for this step height is about 180 nm, for the reasons described previously. During the hard-pad polish step, the reflectance of the wafer surface is monitored with a red laser that shines through a transparent window integrated into the pad. A user-defined wafer average metric of the reflectance is plotted as a function of the polish time as shown in figure 4. When a user-selected criterion is reached (as defined by the pair of blue boxes in figure 4), the CMP tool declares the process end-pointed, and moves on to an overpolish step of a user-defined duration. The endpoint criterion is usually developed iteratively, and includes testing with multiple patterned wafers to ensure it is robust in 'catching' the endpoint. In this process, the endpoint criterion is that the slope of the reflectance curve exceeds a target value twice (for robustness). After the endpoint is triggered (the dotted vertical line in figure 4), the process is continued for 40 seconds of overpolish step.

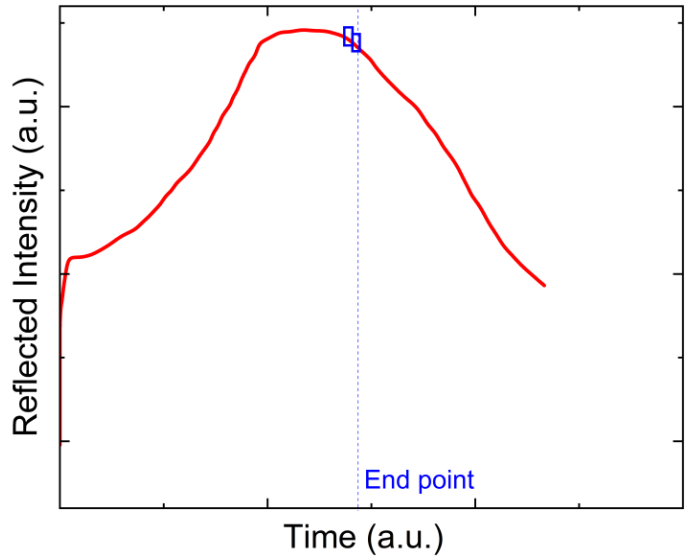

FIG. 4. End point (EP) algorithm trace used to determine a suitable point to declare the completion of hard-pad based removal of Ta.

A test wafer that went through the endpoint followed by the overpolish was cross-sectioned and imaged with SEM to verify the wafer state. As shown in figure 5(a) (schematic cross-section) and figure 5(b) (SEM cross-section), the Ta surface has been locally planarized, leaving a thickness of ~40 nm on the field. The hard-pad process leaves microscratches on the surface, and hence it is desirable to leave some overburden on the field at the end of the hard-pad process, allowing microscratches to be buffed away by the subsequent soft-pad polish step, with minimal removal of silicon oxide in the field areas.

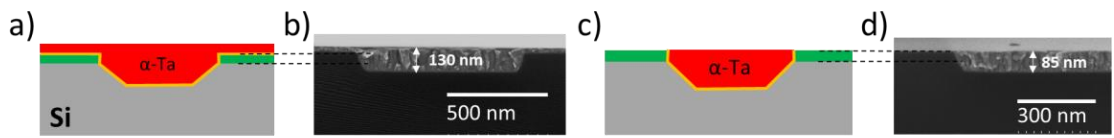

FIG. 5. (a) Schematic showing wafer cross-section after hard pad CMP process, (b) SEM cross section showing the wafer after hard pad CMP process, (c) Schematic showing wafer cross-section after soft pad CMP process, (d) SEM cross section showing the wafer.

Figure 5(c) and 5(d) shows the final schematic and SEM cross section after both steps of the CMP process are completed. It can be seen that the loss of oxide in the field areas is as low as 5 - 10 nm (starting thickness of oxide in the field area is ~50 nm, and the thickness is ~43 to 45 nm when measured by spectroscopic ellipsometry post CMP).

## *4. X-ray fluorescence for patterned wafer CMP process development*

CMP rates on blanket wafers do not translate in a straightforward fashion to rates on patterned wafers. In order to accelerate process development on patterned wafers, X-ray fluorescence (XRF) spectroscopy was used. In particular, we used energy dispersive XRF equipped with a Mo Kα X-ray source with pattern recognition capability to monitor the Ta $L_{\alpha 1}$ signal at ~ 8.15 keV (Ta XRF). For wafers that were processed through both the hard pad and soft pad steps, the trench pattern is revealed. In this case, we added optical pattern-recognition step in the XRF measurement to land on the targeted measurement area.

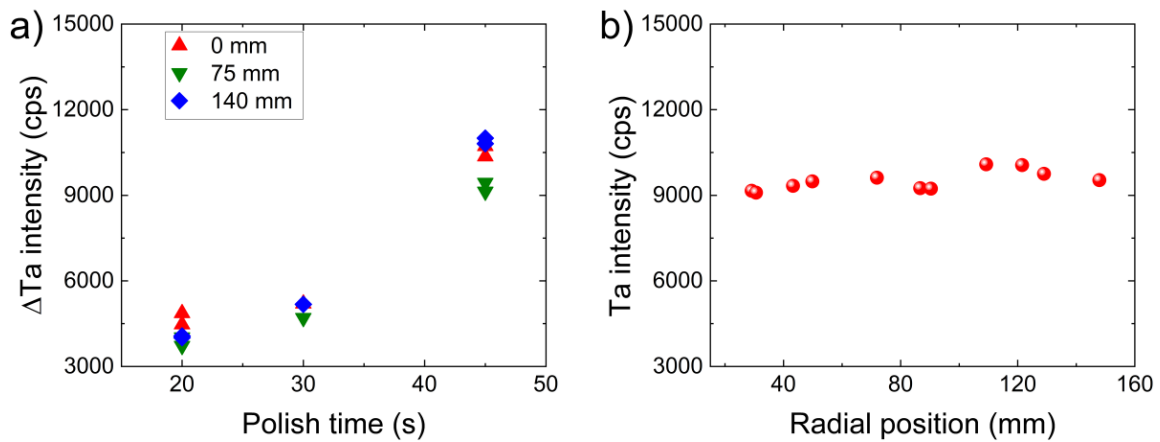

FIG. 6. (a) Change in Ta XRF signal intensity with different soft pad polish times at various wafer radii. (b) Ta XRF signal intensity from a wafer that was processed through optimized CMP process developed in this work.

A 500 μm region with dummy fill was measured by XRF. Figure 6(a) shows a decrease in Ta XRF signal intensity as polishing time increases, reflecting the removal of Ta atoms as CMP progresses. Each point on the graph represents the average of four measurements taken at different (x, y) co-ordinates but with the same wafer radius. Figure 6(b) shows the Ta XRF intensity (after both hard pad and soft pad CMP steps) as a function of radial position from a wafer with 200 nm thick Ta (pre CMP). The data indicates < 4% variation over the entire patterned wafer with the optimized CMP process developed in this work.

## 5. Features due to sub-optimal CMP

This section details how features on the wafer can be studied to reveal sub-optimal CMP processes. The examples presented in this section are derived from earlier iterations of process development, or deliberate deviations from optimal conditions that were tested to determine the process window. Figure 7 shows one of the examples where there are metal residues on field when the polish parameters are not optimized. Figure 7(a) shows the optical microscope image of the patterned feature with dummy metal fill around the patterned features. It can be seen that the metal residues are present on the dummy metal region which is due to non-optimal CMP process parameters. Figure 7(b) shows the SEM micrograph of an example where the Ta metal residues are present even on the patterned feature and figure 7(c) shows a higher magnification view of the Ta metal residue present on the patterned feature, all showing the importance of establishing an optimized CMP process.

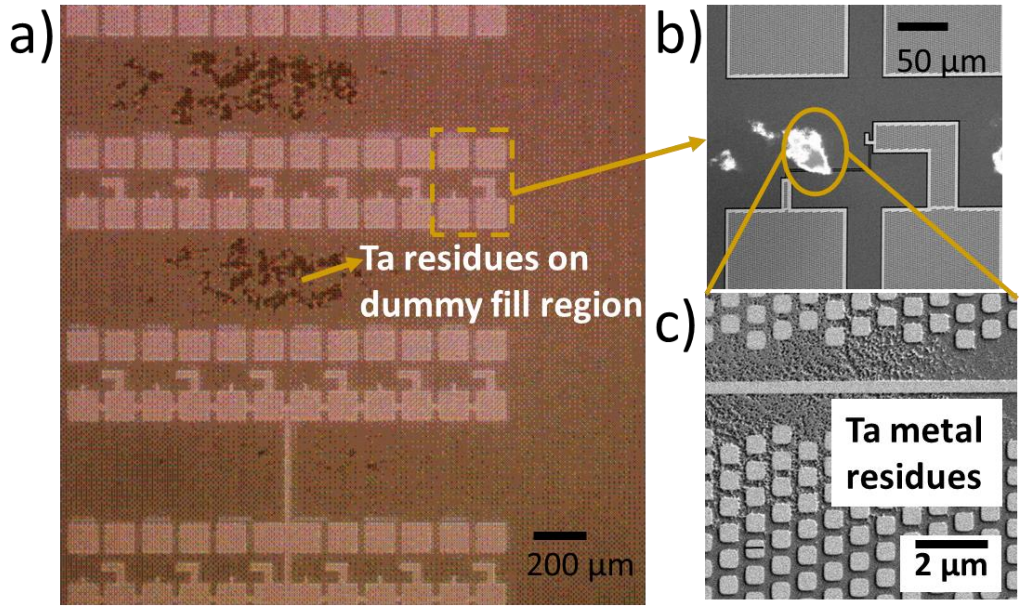

FIG. 7. (a) An example showing the optical microscope image when the CMP process was not optimized properly; CMP residues are seen in the dummy fill area. (b) SEM micrograph showing CMP metal residues on the patterned feature. (c) Higher magnification view of the metal residues in SEM micrograph.

## *6. Influence of the incoming Ta film thickness on CMP*

As described previously, deposited film thickness is usually twice the height difference that needs to be planarized. Figure 8 shows a case when the deposited metal thickness was deliberately chosen to be thinner, at 150 nm rather than 200 nm. Figure 8(a) - (h) shows the comparison of patterned features (post CMP) at wafer center and wafer edge for the two different thicknesses of Ta. Here, we observe more Ta metal residues for the wafer with 150 nm thick Ta whereas the features are clean for the wafer with 200 nm thick Ta for the same CMP process. It can also be noted that the wafer edge is more prone to such residue defects for the chosen CMP process.

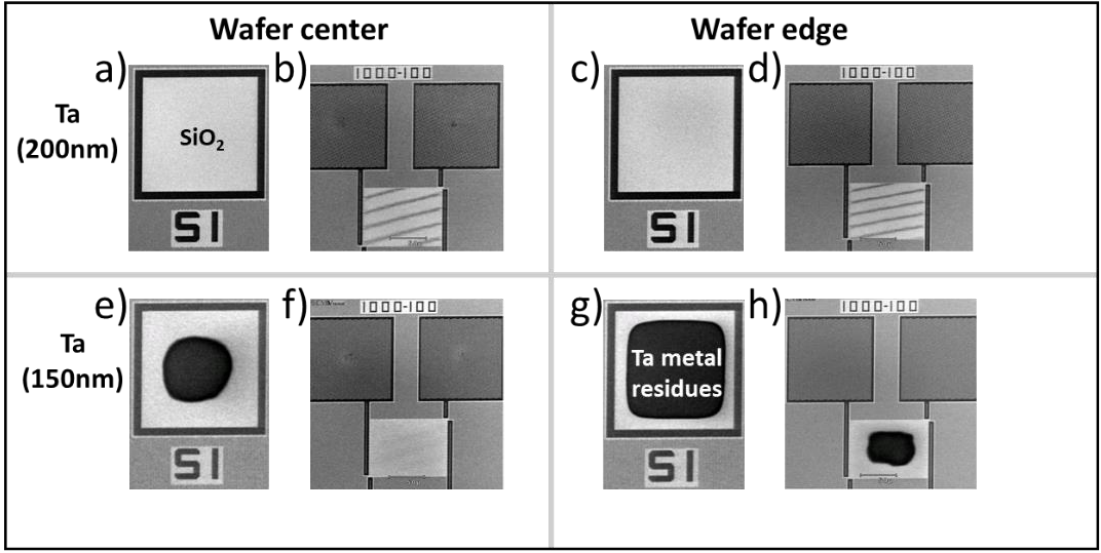

FIG. 8. SEM micrographs showing the different patterned features on wafer at wafer center and wafer edge for different Ta thicknesses used for CMP. (a) through (d) show features on a wafer using 200 nm thick Ta (prior to CMP), at wafer center and wafer edge; (e) through (h) show the corresponding features on a wafer using 150 nm Ta (prior to CMP); Ta residues are visible only on the wafer using 150 nm Ta (prior to CMP).

## III. RESULTS AND DISCUSSION

Several electrical and physical properties need to be well characterized, and produced in a repeatable and predictable fashion, for a designer to use Ta lines in superconducting quantum or superconducting logic circuits. In addition to properties related to superconductivity (transition temperature, coherence length, penetration depth, and critical current density), it is also important to provide the designer with the resistance per square of various line-widths, along with maximum expected range in this resistance due to within-wafer and wafer-to-wafer variation, as well as due to local metal density variations.. Mutual and self-inductance values are usually predictable using geometry-based simulations, and also must be provided to the designer. For superconducting circuits that operate at cryogenic temperatures, many low-level leakage mechanisms (such as conduction through silicon) that are operative at room temperature are absent at operating conditions when thermally excited charge carriers are 'frozen' out. Hence leakage measurements at room temperature are required to assure the designer that shorts associated with residual metal are never present, or particulate contaminants do not accidentally connect two superconducting lines on the chip. In the following sub-sections, the results of our process development are discussed, with reference to resistance and leakage behavior, in addition to establishment of dummy-block and cheese-block rules.

## A. Resistance characteristics of damascene Ta lines

Figure 9 shows the sheet resistance measurements of Ta serpent lines as a function of line width, line spacing, and metal density. Figure 9(a) shows the wafer map with the squares representing individual dies (chips) of dimensions 12.5 mm x 15.5 mm. The squares marked by 'x' represent the 26 sites where the electrical characteristics were measured. It can be seen from figure 9(b) that the sheet resistance of Ta lines remains close to 2 ohms/square for line widths (LW) varying from 100 nm to 3 μm and line spacing (LS) varying from 100 nm to 1 μm. The within-wafer non-uniformity (standard deviation, expressed as a % of the mean value) was determined from measurements taken from 26 dies distributed across the wafer (Figure 9(a)) for each isolated line width with 2 measurements taken from each die for the same copy of device but at different location on the die, as shown in figure 9(c). It can be seen that the sheet resistance non-uniformity is less than 12% for all line widths. The improvement of uniformity as line-width increases can be explained by invoking a fixed line width variation post RIE of 5 nm (independent of line-width). This gives rise to a decreasing % variation as line-width increases. This decreasing %NU trend is added to a CMP-related non-uniformity of ~ 5% that is line-width independent. Figure 9(d) shows the sheet resistance as a function of % metal density. The sheet resistance remains close to 2 ohms/square for all metal pattern densities up to >90%.

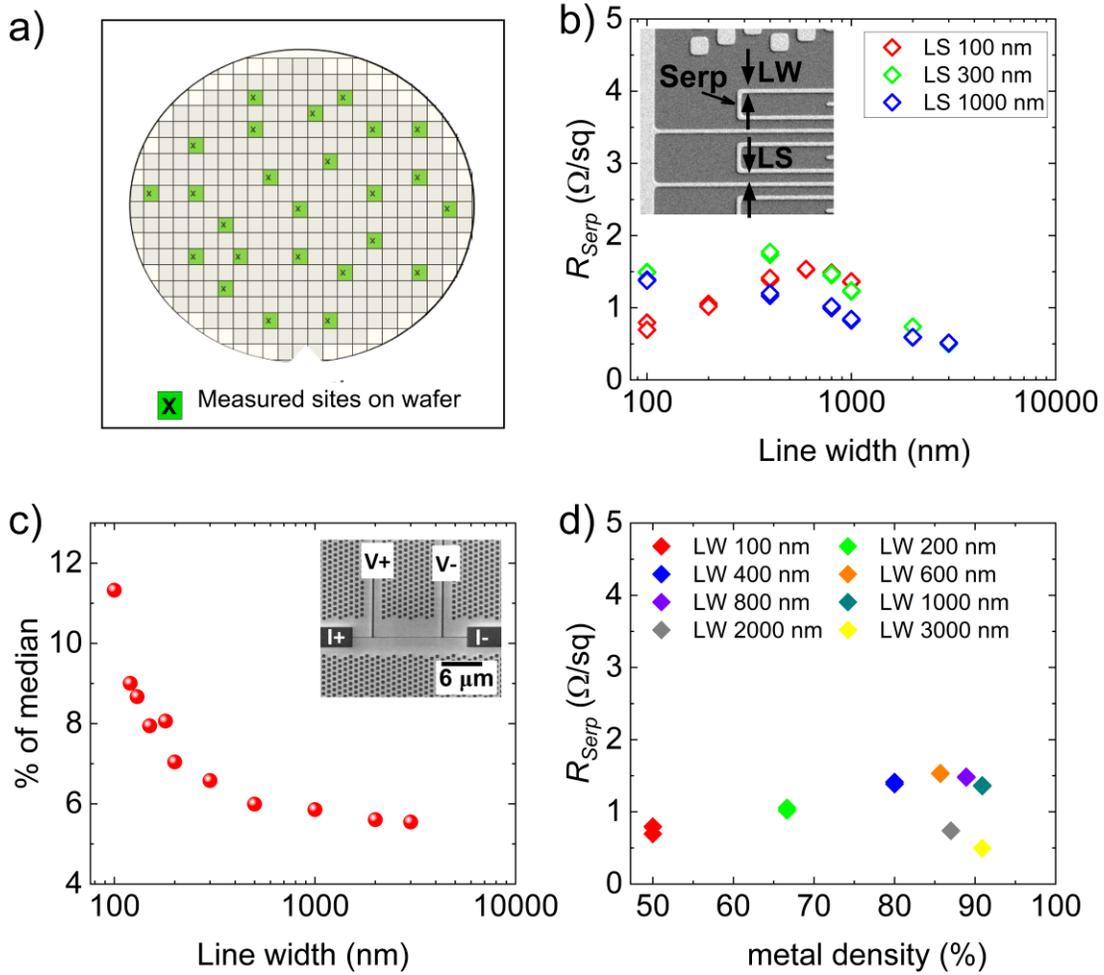

FIG. 9. Electrical test results at room-temperature: a) Wafer map showing the 26 measurement sites for the electrical tests. b) Serpent resistance as a function of line width for line spacing varying from 100 nm to 1000 nm. Inset shows the serpent between comb lines. c) With-in-wafer non-uniformity for isolated lines with widths varying from 100 nm to 3000 nm. Inset shows the 100 nm wide isolated line and the four probe contacts. d) Serpent resistance as a function of metal density for all the line widths as a function of pattern density.

## B. Leakage characteristics between unrelated Ta lines:

As described previously, room temperature leakage measurements of superconducting circuits serve a different purpose than such measurements would have for CMOS integrated circuits operating at room temperature. Leakage measurements in this case are designed to confirm that metallic residues (say due to underpolish) are not present on the wafer. They also serve to confirm that particulate defects that can short two adjacent lines at room temperature do not exist. Such defects cause the leakage to rise to the current compliance limit chosen for room temperature testing. It is also useful to confirm that the measured leakage for a given structure is largely invariant across the wafer diameter, and from wafer to wafer, as another indication of process control.

Figure 10 shows the room temperature leakage characteristics of comb-serpent structures for line space varying from 100 nm to 1 μm and line widths varying from 100 nm to 3 μm. Figure 10(a) shows the SEM micrograph of comb-serpent structure of dimensions 75 μm x 100 μm across which the leakage is measured. Figure 10(b) shows a higher magnification view of the same comb-serpent structure. In contrast, Figures 10(c) and 10(d) show an example of the comb-serpent structure with residual Ta left when the CMP process is not optimal. Such defects will result in excessive leakage currents being measured, triggering a compliance limit during test, as shown in Figures 10(e) and 10(f) (expressed as nA/μm). Leakage values of less than 1 nA/μm were observed for all the line spacings tested at all the points on the wafer, as shown in figure 10(e). Figure 10(f) shows leakage as a function of line spacing for different line widths. While the leakage is low for all the structures shown, it is not surprising to note that the leakage is higher for structures with smaller line space. When the applied voltage is the same for all tested structures, smaller spacing results in a higher electric field, resulting in higher leakage due to charge carrier transport at room temperature.

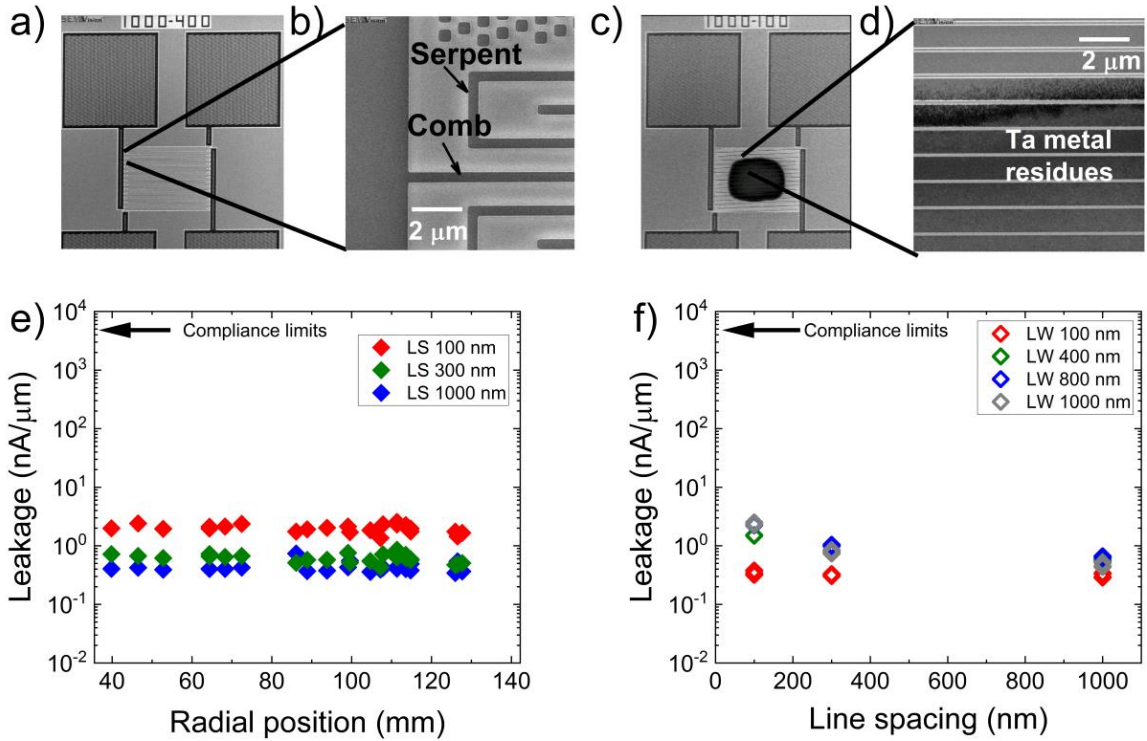

FIG. 10. Room temperature leakage characteristics of comb-serpent structures after CMP. SEM micrographs showing (a) top-view and (b) zoomed-view of comb-serpent structures. (c) top-view and (d) zoomed-view of comb-serpent structure having metal residues due to sub-optimal CMP. (e) Leakage as a function of wafer radius for line spacing varying from 100 nm to 1000 nm, in all cases for a line width of 100 nm. (f) Leakage as a function of line spacing for line widths varying from 100 nm to 3000 nm, for a given die.

## *C.  Local Topography*

Figure 11 shows the results of local topography measurement, using AFM, of a 2 µm wide line after the CMP was completed. Topography is less than 1 nm over a 2 µm wide line, as shown in figure 11(b), for a region marked in blue dashed line in figure 11(a). The second trace indicates the topography across the 15 µm width of the scanned area including the dummy fill region, as shown by green dashed line in figure 11(a). It can be seen that the dummy fill pattern protrudes less than 1 nm from the oxide field area.

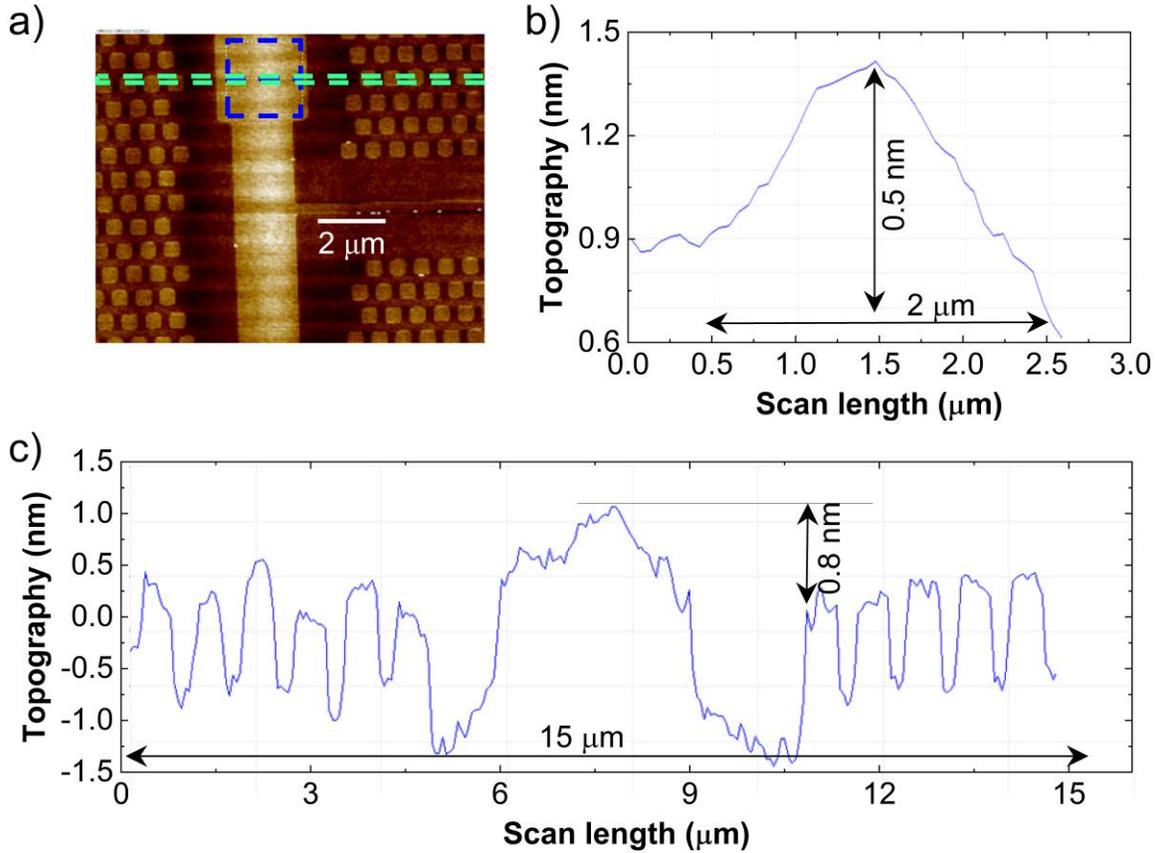

FIG. 11. AFM measurement showing post-CMP topography. (a) Optical picture of the top view of 2 μm wide line with dummy fill around. (b) Measurement showing a step height of <1 nm over a large feature as outlined by a dashed blue box (>2μm wide) in (a). (c) Trace across entire scan width as outlined by a dashed blue box (>2μm wide) in (a).

## *D. Dishing measurement*

Figure 12 shows the topography across a 35 μm wide AFM test feature (inset of figure 12(a)) having line widths varying from 100 nm to 5 μm. The onset of CMP dishing depends on the geometry of the pattern such as the line width and the area fraction of Ta. CMP dishing of less than 2 nm was observed for each line in the AFM test feature at both wafer center (figure 12(a)) and wafer edge (figure 12(b))). Overall topography is less than 5 nm (excluding noise in the AFM scan data) – this bodes well for building multiple layers of Ta interconnect in an oxide matrix for SFQ circuits.

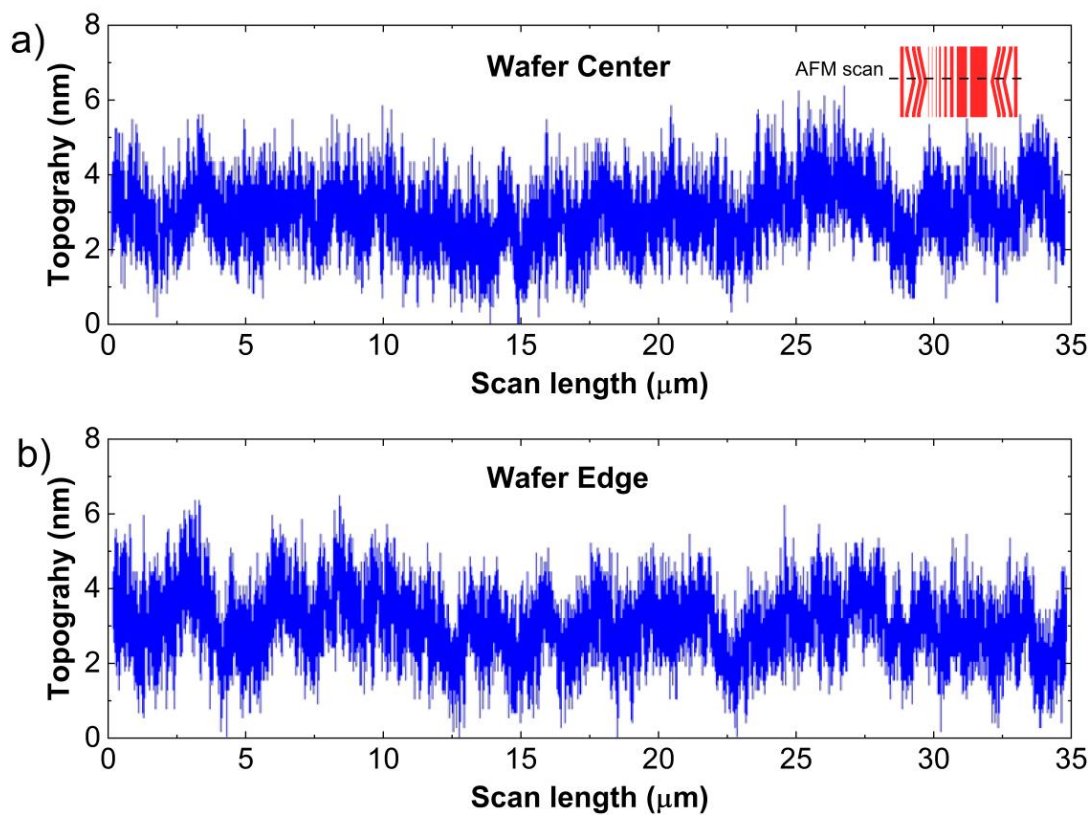

FIG. 12. (a) AFM scan across the black dashed line shown in inset for wafer center. (b) AFM scan across the same black dashed line for wafer edge. The inset shows the structure across which the AFM scans are made. The red color in the AFM structure is Ta and the white spaces between red lines is the $SiO_2$.

## E. CMP Dummy Fill and Cheesing

In regions with no metal traces placed by the designer, it is necessary to add non-functional patterns (called 'dummy fill'), in order to have a uniform metal density across the die (and wafer) [6]. We have chosen a dummy fill unit cell composed of 500 nm square Ta islands, with offsets, as shown in schematic figure 13(a), and filled the areas with no circuit-patterns with dummy fill, as shown in SEM image of figure 13(b). The designer would need to know how close the dummy fill comes to patterned lines. In our design rule manual, this value is set to 6 µm. Additionally, in some regions of the circuit layout, it may be necessary to disallow dummy fill. This is accomplished by a separate dummy block layer (DBLK layer) in the chip design layout file where polygons are placed on areas where automated dummy fill is to be excluded. Residual metal could change the flux distribution and the resonant frequency of superconducting microwave resonators and superconducting qubits, and hence a robust CMP process with an associated design rule that eliminates extraneous residual metal is necessary.

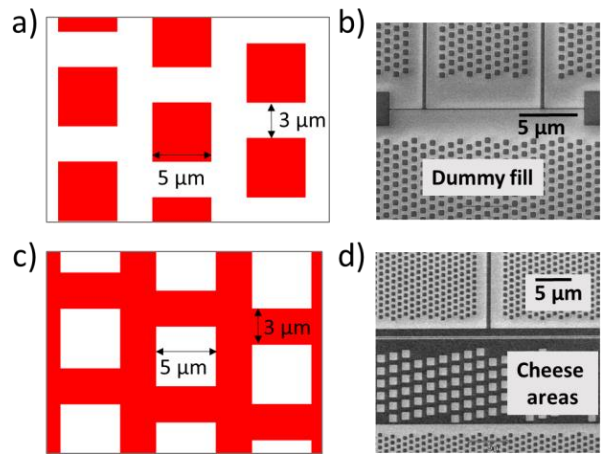

FIG. 13. a) Dummy fill layout image, red color represents the Ta and white color represents absence of Ta; b) Dummy fill SEM micrograph around the Ta patterned features after CMP; c) Cheesing areas layout image; d) Cheesing areas SEM micrograph in the Ta pad after CMP.

In a fashion analogous but opposite to dummy fill, solid metal lines laid out by the designer (that are wider than a certain value), are subjected to automated routines that create holes in a pre-specified pattern. With such 'cheesing' of wide metal lines, the local metal density is reduced from 100% to a value closer to 50%. Wide metal lines were cheesed with holes of dimensions 500 nm x 500 nm, in a pattern as shown in figure 13(c) with the SEM image of the cheesed line shown in figure 13(d). The designer would also need to know how close to the edge of the circuit trace the cheesing holes are allowed to come, and the resultant impact on line resistance. Experimental structures to determine these values are described in the next section.

## F.  Dummy Block Limiting Size

Figure 14 shows the patterned structures with dummy block areas having dimensions varying from 6 µm$^2$ to 150 µm$^2$. The regions with dummy block area up to 75 µm$^2$ have clean features (figure 14(a), (c)) whereas the regions with 150 µm$^2$ area have residual Ta metal leftover, as shown in figure 14(e). Figure 14(b), (d), (f) are the corresponding layout images. This data supports a design rule that dummy block area cannot be larger than 75 µm$^2$.

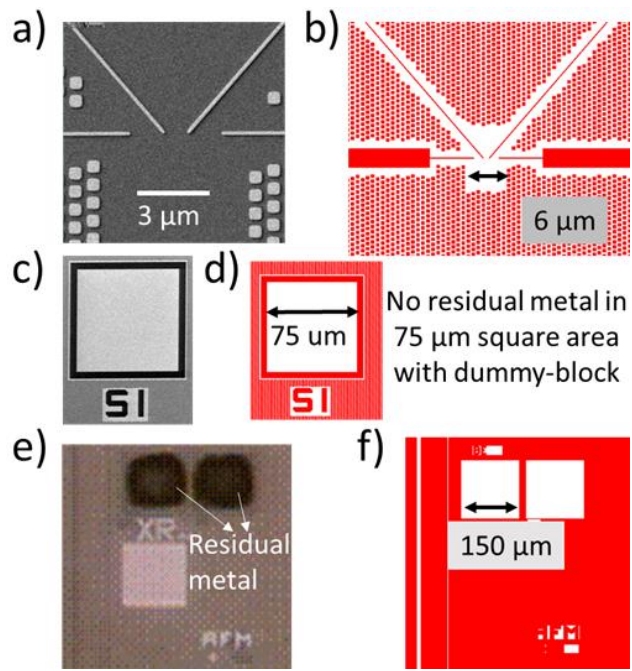

FIG. 14. Effect of dummy block area: (a), (c), (e) are SEM micrographs showing dummy block areas of 6 µm$^2$ to 150 µm$^2$; (b), (d), (f) are the corresponding layout images. Red color presents Ta and white spaces present SiO$_2$ in layout images.

## G. Electrical Impact of Cheese Block Design

The effect of cheesing is measured using electrical measurements in the structure labeled "CMD" as shown in Figure 15(a). In this structure, the width of the uncheesed Ta line (cheese block) is systematically varied. Images from the layout file are shown in figure 15 (b) and figure 15(c) for two different widths of cheese block. Immediately next to the uncheesed Ta line (and separated by 110 nm space), a thin line (110 nm) was placed for 2-point resistance measurement, as shown in SEM image of figure 15(c). This 110 nm wide Ta line is used as a 'canary' to detect the resistance change due to cheese-block.

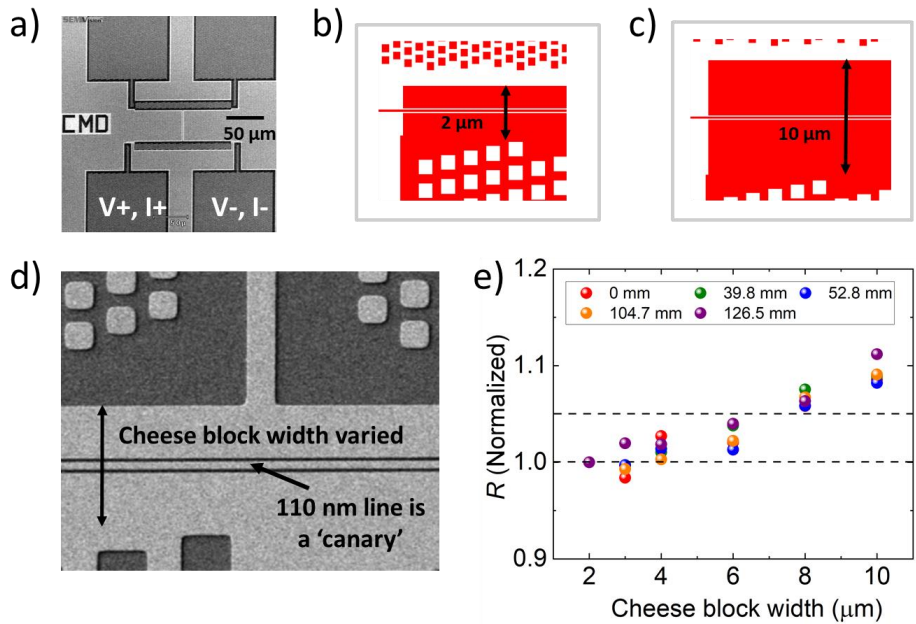

FIG. 15 (a) SEM micrograph showing the two probe measurement arrangement for 110 nm wide line; (b) and (c) shows the layout images for two different widths of cheese block layer; (d) SEM micrograph showing the 110 nm wide line used to measure the effect of cheesing; (e) Normalized resistance as a function of cheese block width at various radial positions on the wafer (as labeled). The dotted line at a normalized resistance of 1.05 indicates the 5% criterion used to set the maximum permitted cheese block width.

Figure 15(e) is a plot of the normalized resistance as a function of cheese block width at wafer center, and wafer radii of 52.79 mm, 104.69 mm, and 126.52 mm. The normalization is with respect to the resistance at 2 µm cheese-block width. The resistance increased by ~8.5% as the cheese-block width was increased from 2 µm to 10 µm for all radii. In order to keep the resistance change to < 5%, the limiting width of cheese-block has been set at 6 µm.

### *H.  Wafer to Wafer Repeatability*

The optimized CMP recipe was subsequently tested on four wafers over a period of two months. The isolated line sheet resistance results were repeatable, with all wafers showing the same trend as a function of linewidth, as shown in figure 16(a). Figure 16(b) shows that the leakage (for a line spacing of 300 nm) was low at all the measured sites on all wafers. Figure 16(c) compares the sheet resistance for a 1 µm wide serpent (spaced 300 nm away from unrelated metal lines) for these wafers, providing an illustration of the degree of repeatability observed with this process. The influence of dummy block and cheese block on CMP performance was also the same in terms of maximum allowed dimensions (75 µm$^2$) with clean features and the impact of cheese block width on the Ta line resistance, normalized to the resistance with 2 µm cheese-block width (Figure 16(d)).

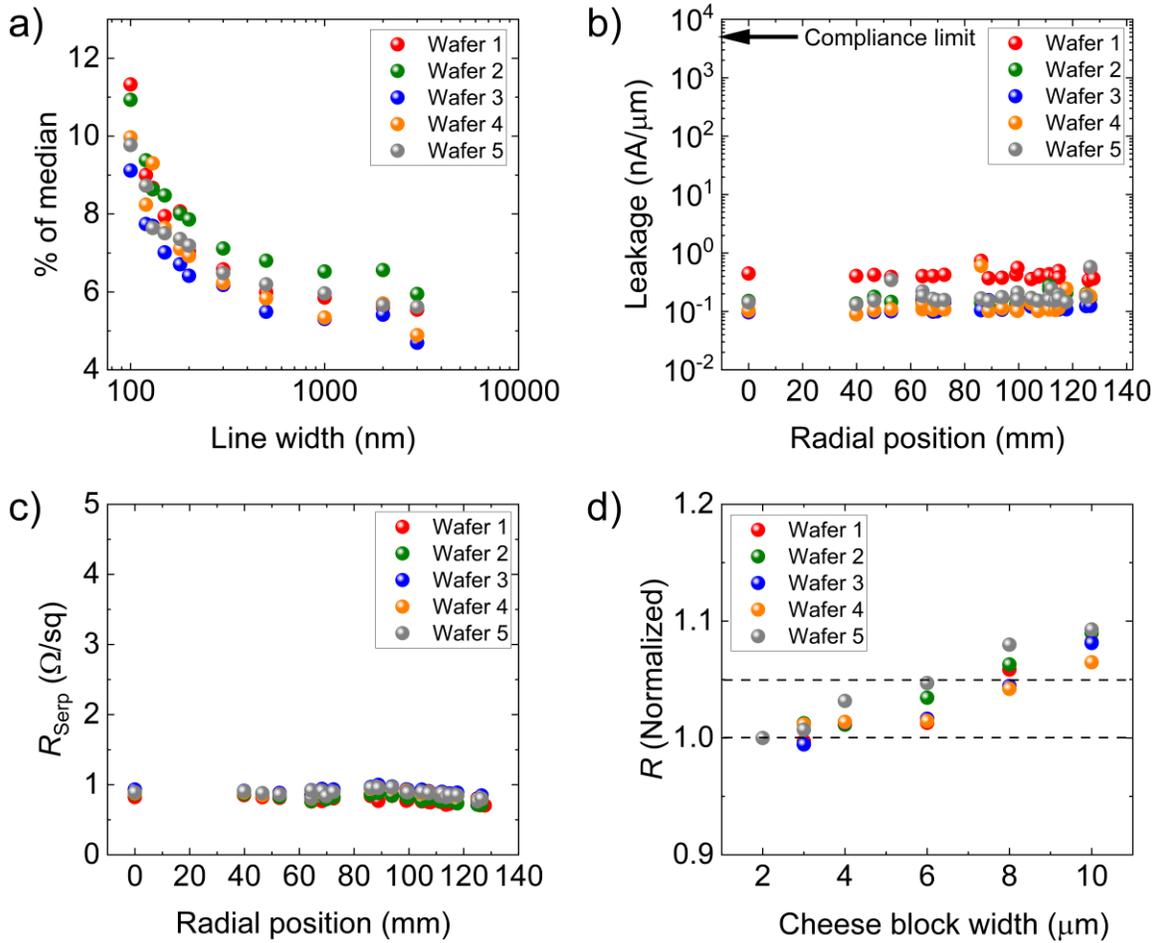

FIG. 16. Repeatability test showing a comparison of (a) With-in-wafer non-uniformity of sheet resistance as a function of line width, (b) Leakage as a function of wafer radius for a line spacing of 300 nm, (c) Serpent sheet resistance as a function of wafer radius for a line width of 1000 nm and line spacing of 300 nm, (d) Ta line resistance (normalized) as a function of cheese block width.

## IV. CONCLUSIONS AND FUTURE WORK

We developed a process flow for creating damascene α-Ta patterns (~85 nm thick) for the first time on 300 mm wafer scale. The results presented here form part of a 300 mm Ta CMP design-rule manual that specifies minimum linewidth and space, maximum dummy block area, maximum uncheesed line width and provides the electrical characteristics of Ta lines of line-widths ranging from 100 nm to 3 µm. The process described here can be utilized for making superconducting resonators, and for superconducting interposer chips for connecting multiple chips. While the process was tested in a single layer, the topography measurements post CMP suggest that SFQ circuits with multiple layers can utilize these results. The CMP process employed an optical endpointing during the hard pad polish step, followed by a fixed time overpolish on a soft pad. This process was tested over a period of two months to test the repeatability. Dummy block regions with dimensions up to 75 µm$^2$ were found to be safe, without any residual metal observed. Maximum cheese block width was set to 6 µm. Leakage value of less than 1 nA/µm is reported for a 50% metal density. Post CMP topography was determined to be less than 5 nm at all points on the wafer for all the line widths tested. Serpent sheet resistance was found to range from a minimum of 0.44 ohms/square for 3 µm lines to 1.9 ohms/square for lines of 400 nm width. Within-wafer resistance non-uniformity was determined to be less than 12% for all line widths. In future studies, the post RIE CD uniformity will be improved. During CMP processing, a first wafer effect was observed – the resultant increase in variability can be addressed by using one 'seasoning wafer' ahead of the product wafers. In addition, it is likely that long-term repeatability of the post-CMP resistance performance can be further improved by implementing a 'rate-qualification' procedure and replacing the fixed time polish used in this work with a polish time that takes into account rate decay due to pad aging.

# ACKNOWLEDGMENTS
The support of this work by the Air Force Research Lab (AFRL), Rome, NY, through Contract FA8750-19-1-0031, and Contract FA864921P0773, is gratefully acknowledged. The authors gratefully acknowledge the help provided by Matthew Bracken of NY CREATES for cross-sectional analyses by scanning electron microscopy.# AUTHOR DECLARATIONS

**Conflicts of Interest** *(required)*

The authors have no conflicts to disclose.

# DATA AVAILABILITY

The data that support the findings of this study are available from the corresponding author upon reasonable request.